\documentclass[12pt]{article}
\input{epsf.sty}

\def\mev{\,{\rm Me\kern-0.1em V}}
\def\gev{\,{\rm Ge\kern-0.1em V}}

\usepackage{color}
\usepackage[dvips]{graphicx} 
\usepackage{colordvi} 
\begin{document}
\vspace*{-1.25in}
\small{
\begin{flushright}
FERMILAB-Pub-03/197-T \\[-.1in]
%July~2003 \\
\end{flushright}}
\vspace*{.75in}
\begin{center}
{\Large{\bf  Chiral Lagrangian Parameters
for Scalar and Pseudoscalar Mesons}}\\
\vspace*{.45in}
{\large{W. ~Bardeen$^1$,
E.~Eichten$^1$, and
H.~Thacker$^2$}} \\ 
\vspace*{.15in}
$^1$Fermilab, P.O. Box 500, Batavia, IL 60510 \\
$^2$Dept.of Physics, University of Virginia, Charlottesville, 
VA 22901
\end{center}
%%%%%%%%%%%%%%%%%%%%%%%%%%%%%%%
\vspace*{.2in}
\begin{abstract}
The results of a high-statistics study of scalar and pseudoscalar meson propagators in
quenched lattice QCD are presented. For two values of lattice spacing, 
$\beta=5.7$ ($a \approx .18$ fm) and $5.9$ ($a \approx .12$ fm), 
we probe the light quark mass region using clover improved 
Wilson fermions with the MQA pole-shifting ansatz to treat the 
exceptional configuration problem. 
The quenched chiral loop parameters $m_0$ and $\alpha_{\Phi}$ 
are determined from a study of the pseudoscalar hairpin correlator. 
From a global fit to the meson correlators, estimates are obtained
for the relevant chiral Lagrangian parameters, including the Leutwyler parameters 
$L_5$ and $L_8$. Using the parameters obtained from the singlet and nonsinglet 
pseudoscalar correlators, the quenched chiral loop effect in the nonsinglet 
scalar meson correlator is studied.
By removing this QCL effect from the lattice correlator, we obtain the mass and 
decay constant of the ground state scalar, isovector meson $a_0$. 
\end{abstract}

\section{Introduction}

Improved methods for studying the regime of small quark mass in lattice QCD 
provide the realistic prospect of quantitatively determining the parameters of  
the low energy chiral Lagrangian of QCD from first principles. Although a 
definitive comparison with experiment requires the analysis of full and/or 
partially quenched simulations, detailed studies of chiral behavior in the 
quenched approximation are of interest for several reasons. First, the 
characteristic quenched chiral loop effects which arise from the anomalous 
double-pole structure of the quenched, flavor-singlet pseudoscalar propagator 
will also occur in partially quenched calculations (at a level determined
by the mismatch between valence and sea quark masses)\cite{SharpeShoresh}. 
The observation of these anomalous effects in the quenched theory should 
provide a useful baseline for future chiral analysis of full QCD. Second, 
although it is not a unitary theory, the quenched approximation can be 
analyzed in an effective Lagrangian framework\cite{B&G,Sharpe}, yielding a 
well-defined set of low-energy constants in quenched chiral perturbation
theory. (In practice, this requires the assumption that the $U_A(1)$ breaking 
from the anomaly can also be treated perturbatively).  Comparison of these 
constants with those of full QCD can provide valuable insight into the role 
of closed quark loops in hadron phenomenology. In addition, the study of chiral
behavior in the quenched approximation provides useful information about 
the interplay between topological charge and chiral symmetry breaking in QCD. 
For example, the Witten-Veneziano formula relates the gluonic component of the 
$\eta'$ mass in full QCD to the topological susceptibility of the quenched theory. 
 
In two previous papers \cite{chlogsI,scalar} we reported results of a study of 
the chiral behavior of scalar and pseudoscalar meson propagators in quenched QCD 
at $\beta = 5.7$ using clover improved Wilson fermions. In this paper we present 
results from a new data set at $\beta=5.9$ \cite{lat02}, compare them with the 
$\beta = 5.7$ results, and summarize the main conclusions of this study. We 
also compare our results to those of other recent studies.\cite{Wittig}
An important ingredient in our analysis is the use of the MQA pole-shifting 
procedure \cite{MQA} to resolve the exceptional configuration problem. Our 
experience with this technique has led us to conclude that it provides a 
practical and quantitatively acceptable resolution of the problem, which 
eliminates the spurious statistical fluctuations of exceptional configurations 
without systematically biasing the final results. This conclusion is
based on the consistency between the light-quark results and those of heavier 
quarks where the pole-shifting has a negligible effect on the propagators. 
It is also supported by the overall agreement we observe between our results 
(in and out of the pole region) and theoretical expectations based on quenched 
chiral perturbation theory. (See Section 6)

Since the introduction
of the MQA procedure, other methods for avoiding the exceptional configuration problem 
have been explored. These include twisted-mass QCD \cite{twisted} and the use of exactly
chiral (overlap \cite{overlap} or domain-wall \cite{Kaplan}) fermions. All these approaches
have in common the fact that Wilson-Dirac eigenvalues at positive real quark mass are eliminated, thus resolving the problem. It should be noted that
exactly real eigenmodes of the Wilson-Dirac operator, which are the cause of the exceptional configuration problem, make a negligible contribution to physical quantities in the 
infinite volume limit (vanishing like $1/\sqrt{V}$). Thus any prescription which effectively removes these poles from the physical region should provide a satisfactory resolution
of the problem for sufficiently large volume. The MQA pole shifting
is a minimal prescription for accomplishing this. A more stringent test 
of the procedure is the study of chiral behavior for very light quarks 
in a {\it finite} volume which is large 
compared to the QCD scale but comparable to the chiral scale. In this regime, finite volume
effects are large but calculable in $Q\chi PT$, simply by replacing loop integrals by
finite-volume momentum sums. As discussed in \cite{scalar}, the scalar, isovector (valence)
meson propagator exhibits a prominent quenched chiral loop effect arising from the
$\eta'$-$\pi$ intermediate state. For the lightest quark masses we study, the finite volume effects expected from $Q\chi PT$ are quite large. Thus, the detailed agreement (as a function of both time and pion mass) between the measured scalar 
propagator and the {\it finite-volume} one-loop calculation provides a convincing demonstration that
the MQA procedure is an effective method for exploring the light quark regime with Wilson
fermions.  

%%%%%%%%%%%%%%%%%%%%%%%%%%%%%%%%%%%%%%%%%%%%%%%%%%%%

\section{Quenched Chiral Perturbation Theory}

To analyze the quenched theory in a chiral Lagrangian framework, one introduces
wrong-statistics ghost quark fields to cancel closed loops, yielding a low-energy 
chiral Lagrangian with a graded $U(3|3)\times U(3|3)$ symmetry \cite{B&G}. 
At the one-loop level, this is equivalent
to the simpler and more direct approach to quenched $\chi PT$ \cite{Sharpe}
which begins with an ordinary $U(3)\times U(3)$ chiral Lagrangian describing
a nonet of Goldstone bosons. To leading order, this is
\begin{equation}
{\cal L}_2 = \frac{f^2}{4}\left[Tr(\partial_{\mu}U^{\dag}\partial^{\mu}U)
+ Tr(\chi^{\dag}U+U^{\dag}\chi)\right]
\end{equation}
where $U$ is a $U(3)\times U(3)$ chiral field and $\chi$ is the pseudoscalar mass matrix.
Our analysis also incorporates the following fourth-order terms in the chiral 
Lagrangian \cite{Leut85},
\begin{equation}
{\cal L}_4 = L_5 Tr(\partial_{\mu}U^{\dag}\partial^{\mu}U(\chi^{\dag}U+U^{\dag}\chi))
+ L_8 Tr(\chi^{\dag}U\chi^{\dag}U + U^{\dag}\chi U^{\dag}\chi)
\end{equation}
The effect of the axial U(1) anomaly is introduced
as an explicit symmetry breaking term consisting of a flavor-singlet pseudoscalar
$\eta'$ mass term and a field renormalization,
\begin{equation}
\label{eq:hairpin}
{\cal L}_{hp} = \frac{1}{2}(\alpha_{\Phi}\partial^{\mu}\eta'\partial_{\mu}\eta'
- m_0^2\eta'^2)
\end{equation}
where 
\begin{equation}
\eta' = \frac{f}{2}\left(iTr\ln(U^{\dag})-iTr\ln(U)\right)
\end{equation}
Finally, we will also analyze the scalar, isovector meson propagator, 
which turns out to be well-described
by a combination of a heavy $a_0$ meson and an $\eta'$-$\pi$ loop diagram. Thus, we
incorporate a scalar-isovector meson field, using the formalism of nonlinear chiral
Lagrangians \cite{chlogsI},
\begin{equation}
{\cal L}_{sc} = \frac{1}{4}tr\{D\sigma D\sigma\} - \frac{1}{4}m_s^2 tr\{\sigma\sigma\}
+ f_s tr\{\chi^{\dag}\sqrt{U}\sigma\sqrt{U}+\chi\sqrt{U}^{\dag}\sigma\sqrt{U}^{\dag}\}
\end{equation}
where D is a chirally covariant derivative.
One of our motivations for studying the scalar correlator is 
the expectation of a prominent quenched chiral loop effect from 
the $\eta'$-$\pi$ intermediate state, as discussed in Ref. \cite{scalar}. 
The agreement between the lattice correlator and the one-loop
calculation is very good, particularly for the $\beta=5.9$ 
results, as discussed in Section 6.

To summarize, the low energy chiral Lagrangian used in our analysis is 
\begin{equation}
{\cal L} = {\cal L}_2 + {\cal L}_4 + {\cal L}_{hp} + {\cal L}_{sc}
\end{equation}
In the quenched approximation, we have the supplementary rule that multiple $\eta'$ mass
insertions on a given pseudoscalar
line are excluded. More generally, any $\chi PT$ diagram corresponding
to a quark-line diagram with internal closed loops is discarded \cite{Sharpe}. At
the one-chiral-loop level, these rules are unambiguous, and equivalent to the more systematic
procedure of introducing ghost fields \cite{B&G}.

%%%%%%%%%%%%%%%%%%%%%%%%%%%%%%%%%%%%%%%%%%%%%%%%%%%%

\section{Lattice Parameters}

The calculations discussed in this paper were carried out on the Fermilab ACPMAPS
and on the UVA Linux cluster GARCIA. The two Monte Carlo gauge ensembles analyzed
consisted of 300 configurations at $\beta=5.7$ on a $12^3\times 24$ lattice, and 350 configurations
at $\beta=5.9$ on a $16^3\times 32$ lattice (the Fermilab b and c ensembles). Quark propagators were
calculated with clover improved Wilson action. The clover coefficients used were $C_{sw}=1.57$
for $\beta=5.7$ and $C_{sw}=1.50$ for $\beta=5.9$. For $\beta=5.7$, the quark propagators were 
calculated for $\kappa = .1410,.1415,.1420,.1423,.1425,.1427,$ and .1428, with $\kappa_c = .14329$,
while for $\beta=5.9$, propagators were calculated with $\kappa=.1382,.1385,.1388,.1391,.1394,$ and
.1397, with $\kappa_c = .14013$. In physical units, this corresponds to a range of pion masses
of 275 to 565 MeV for $\beta=5.7$ and 330 to 665 MeV for $\beta=5.9$. Here and elsewhere, we will
quote results in physical units using the rho mass to set the scale. 

An analysis of smeared and local
rho propagators on our ensembles yields $m_{\rho}a=.690(8)$ and $.469(3)$ for $\beta=5.7$ and 
$\beta=5.9$, respectively.  
An analysis of the axial-vector meson channel also yields a 
mass for the $a_1$ meson of $1.15(7)$ and $0.77(3)$ for $\beta=5.7$ and 
$\beta=5.9$ respectively. 
Using the rho mass to fix the scale gives $a^{-1}=1.12$ GeV for $\beta=5.7$ and 
$1.64$ GeV for $\beta=5.9$. The resulting physical mass for the $a_1$ 
($1290$ MeV for $\beta=5.7$ and $1260$ MeV for $\beta=5.9$) is
close to the mass of the observed $a_1(1260)$ resonance.

To get some idea of the systematic error associated with choice of scale, 
we will sometimes quote equivalent results using the charmonium 1S-1P splitting scales 
of 1.18 GeV and 1.80 GeV for the two ensembles. 
The MQA pole-shifting procedure \cite{MQA} was applied to all quark propagators. 
For $\beta=5.7$, all poles below $\kappa=.1431$ were located and shifted, 
while for $\beta=5.9$, all poles below $\kappa=.1400$ were shifted.

\section{The Hairpin Insertion, $\eta'$ Mass, and Topological Susceptibility} 

We begin by determining the parameters $m_0$ and $\alpha_{\Phi}$ in the term 
${\cal L}_{hp}$, Eq. (\ref{eq:hairpin}). 
These parameters are extracted from the two-quark-loop (``disconnected'') 
piece of the quenched flavor singlet pseudoscalar correlator,
\begin{equation}
\label{eq:hprop}
\Delta_h(x) = \langle Tr\gamma^5G(x,x) Tr\gamma^5G(0,0)\rangle
\end{equation}
The lowest order $Q\chi PT$ approximation to (\ref{eq:hprop}) is 
the tree graph with a single hairpin insertion between two pion propagators.
In momentum space, this is
\begin{equation}  
\tilde{\Delta}_h(p) = f_P\frac{1}{p^2+m_{\pi}^2}\left(m_0^2+\alpha_{\Phi}p^2\right)
\frac{1}{p^2+m_{\pi}^2}f_P
\end{equation}
where $f_P$ is the pseudoscalar decay constant,
\begin{equation}
f_P = \langle 0|\bar{\psi}\gamma^5\psi|\pi\rangle 
\end{equation}
Fourier transforming over $p_0$ and setting $\vec{p}=0$, we have
\begin{equation}
\label{eq:dipole}
\Delta_h(\vec{p}=0,t) = \frac{f_P^2}{4m_{\pi}^3}\left[C_+ + C_-m_{\pi}t\right]e^{-m_{\pi}t}
+ (t\rightarrow T-t)
\end{equation}
where
\begin{equation}
C_\pm \equiv m_0^2\pm \alpha_{\Phi}m_{\pi}^2
\end{equation}

\begin{figure}
\epsfxsize = 0.80\textwidth
\centerline{\epsfbox{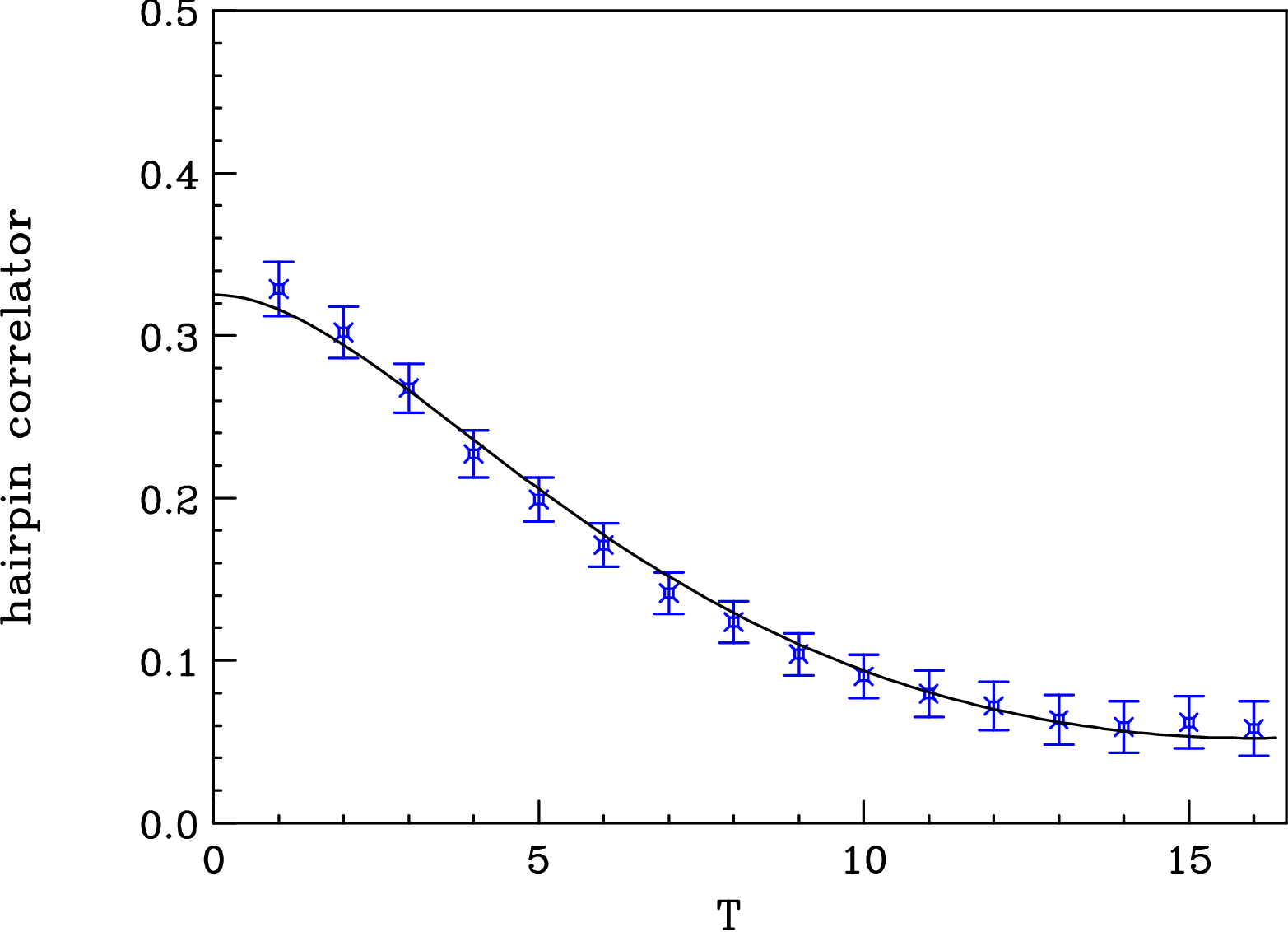}}
\caption[]{The quenched hairpin correlator for $\beta=5.9$, $\kappa=.1394$. 
The solid line is a pure dipole fit with $\alpha_{\Phi}=0$. }
\label{fig:hprop}
\end{figure}

In our previous analysis of the $\beta=5.7$ ensemble, the hairpin
correlator $\Delta_h$ was studied
for both local and smeared sources and compared with the pion pole residues of the 
corresponding valence propagators. This analysis demonstrated a remarkable absence of
excited state contamination in the hairpin correlator, even when the sources were only
separated by one or two time slices. Moreover, the time-dependence of $\Delta_h(t)$ was
well-described at all times $t\geq 2$ 
by the formula (\ref{eq:dipole}) with $\alpha_{\Phi}=0$,
i.e. by a pure momentum-independent mass insertion. For the present analysis, we have
fit both the 5.7 and 5.9 ensembles to the full two-parameter formula (\ref{eq:dipole})
in order to obtain an accurate estimate of $\alpha_{\Phi}$. For the 5.7 ensemble, 
acceptable $\chi^2$'s were obtained by fitting a range of times from $t = 3$ to 12.
Using the fully correlated error matrix, the covariant
$\chi^2$ for these fits ranged from 0.8 to 1.5 per degree of freedom.
For the hairpin correlators at $\beta = 5.9$, 
we obtained very good fits to the formula (\ref{eq:dipole}) over the
entire time range from $t = 1$ to 16. Here the correlated $\chi^2$'s 
ranged from 0.3 to 0.5 per degree of freedom. An example of a hairpin fit
for $\beta=5.9$ and $\kappa=.1394$ is shown in Fig. \ref{fig:hprop}. The solid line is the
pure dipole fit with $\alpha_{\Phi}=0$. In Tables \ref{tab:hairpin1} 
and \ref{tab:hairpin2} we give the $\beta=5.7$ and 5.9
results for $m_0$ and $\alpha_{\Phi}$. Also shown in the last column are the values
of $m_0$ obtained from the 1-parameter pure dipole fit with $\alpha_{\Phi}=0$.
Considering first the results for $\alpha_{\Phi}$, the values for the 5.7
ensemble are negative by about one to two standard deviations, while the values
for $\beta = 5.9$ are slightly positive, also by about two standard deviations.
The values for different $\kappa$'s within each ensemble are highly correlated,
so the deviation of $\alpha_{\Phi}$ from zero in either data set has little 
statistical significance. Ignoring $\kappa$-dependence and averaging the values
within each ensemble, we get
\begin{equation}
\alpha_{\Phi} = -0.15 \pm 0.10, ~~~\beta = 5.7
\end{equation}
and
\begin{equation}
\alpha_{\Phi} = 0.05 \pm 0.03, ~~~\beta = 5.9. 
\end{equation}

If we ignore any possible lattice spacing dependence and
average the two data sets in quadrature, we get the final result
\begin{equation}
\alpha_{\Phi} = 0.03 \pm 0.03
\end{equation}
This can be regarded as a success of the large-$N_c$ view of
the anomaly where this renormalization is an order $1/N_c$ effect.
In the subsequent analysis we will take $\alpha_{\Phi} = 0$.

\begin{table}
\centering
\caption{Fit parameters $m_0$ and $\alpha_{\Phi}$ for the $\beta=5.7$ hairpin correlators. All masses are in lattice units.}
\vspace*{0.5cm}
\label{tab:hairpin1}
\begin{tabular}{||c|c||c|c||c||}
\hline 
$\kappa$ & $m_{\pi}$ & $m_0$ & $\alpha_{\Phi}$ & 
$m_0 (\alpha_{\Phi}=0)$  \\
\hline
.1410 & .505(2)  & .269(26) & -.17(10) & .280(10)  \\
.1415 & .450(3) & .291(24) & -.18(10) & .294(10)  \\
.1420 & .386(3) & .310(21) & -.19(10) & .308(10)  \\
.1423 & .342(4) & .321(19) & -.19(10) & .316(10)  \\
.1425 & .307(4) & .326(19) & -.16(11) & .321(10)  \\
.1427 & .267(5) & .326(19) & -.10(12) & .322(11)  \\
.1428 & .245(6) & .323(19) & -.03(13) & .322(11)  \\
\hline
\end{tabular}
\end{table}

%%%%%%%%%%%%%%%%%%%%%%%%%%%%%%%%%%%%%%%%%%%%%%%%%%%%
\begin{table}
\centering
\caption{Fit parameters $m_0$ and $\alpha_{\Phi}$ for the $\beta=5.9$ hairpin 
correlators. All masses are in lattice units.}
\vspace*{0.5cm}
\label{tab:hairpin2}
\begin{tabular}{||c|c||c|c||c||}
\hline 
$\kappa$ & $m_{\pi}$ & $m_0$ & $\alpha_{\Phi}$ & 
$m_0 (\alpha_{\Phi}=0)$  \\
\hline
.1382 & .411(3) & .188(14) & .04(2) & .194(5)  \\
.1385 & .378(3) & .192(13) & .04(2) & .198(5)  \\
.1388 & .343(4) & .197(11) & .04(2) & .203(5)  \\
.1391 & .304(4) & .204(11) & .05(2) & .209(5)  \\
.1394 & .261(5) & .211(9)  & .06(3) & .217(5)  \\
.1397 & .204(5) & .220(9)  & .11(4) & .226(6)  \\
\hline
\end{tabular}
\end{table}
%%%%%%%%%%%%%%%%%%%%%%%%%%%%%%%%%%%%%%%%%%%%%%%%%%%%

Fitting the hairpin correlators to the pure dipole $\alpha_{\Phi}=0$ form (last column of Tables \ref{tab:hairpin1} 
and \ref{tab:hairpin2}) we obtain the chirally extrapolated values (in lattice units)
\begin{equation}
\label{eq:m01}
m_0 = .348(4), ~~~ \beta = 5.7 
\end{equation}
and 
\begin{equation}
\label{eq:m02}
m_0 = .232(4), ~~~\beta = 5.9. 
\end{equation}
Using the rho mass scale
and including a flavor factor of $\sqrt{3}$, this gives the
gluonic component of the $\eta'$ mass
\begin{eqnarray}
m^{glue}_{\eta'} & = & 675(8)\; {\rm MeV}, ~~~\beta = 5.7 \\
                 & = & 659(12)\; {\rm MeV}, ~~~\beta = 5.9
\end{eqnarray}
If we instead use the charmonium scale we get
\begin{eqnarray}
m^{glue}_{\eta'} & = & 712(9)\; {\rm MeV}, ~~~\beta = 5.7 \\
                 & = & 723(13)\; {\rm MeV}, ~~~\beta = 5.9
\end{eqnarray}
We conclude that the $\eta'$ mass scales reasonably well between $\beta=5.7$ and 5.9,
well within the systematic uncertainty associated with different ways of determining
the lattice spacing. The values we obtain for the $\eta'$ mass insertion are somewhat
low compared to the estimate of $\approx 850$ MeV obtained from the physical 
$\eta'$ mass and chiral perturbation theory. Although we see approximate scaling, it would require calculations at larger values of $\beta$ to rule out a significant
lattice spacing effect. It is also worth remembering
that the whole framework in which the quenched 
hairpin diagram is interpreted as a mass insertion
is only demonstrably valid in the limit of large $N_c$, so some discrepency between the lattice calculation of
$m_0$ and the phenomenological estimate might be expected. 
A recent calculation of the $\eta'$ mass in two-flavor {\it full} QCD by the 
CPPACS collaboration \cite{CPPACS} gave the result 
$m_{\eta'} = 960(87)^{+36}_{-286} {\rm MeV}$, in good agreement with experiment.
Detailed comparisons between quenched and full QCD studies of the $\eta'$ should provide a better understanding
of the accuracy of large-$N_c$ arguments in the framework of chiral Lagrangians.

\begin{figure}[p]
\epsfxsize = 0.90\textwidth                                                     \centerline{\epsfbox{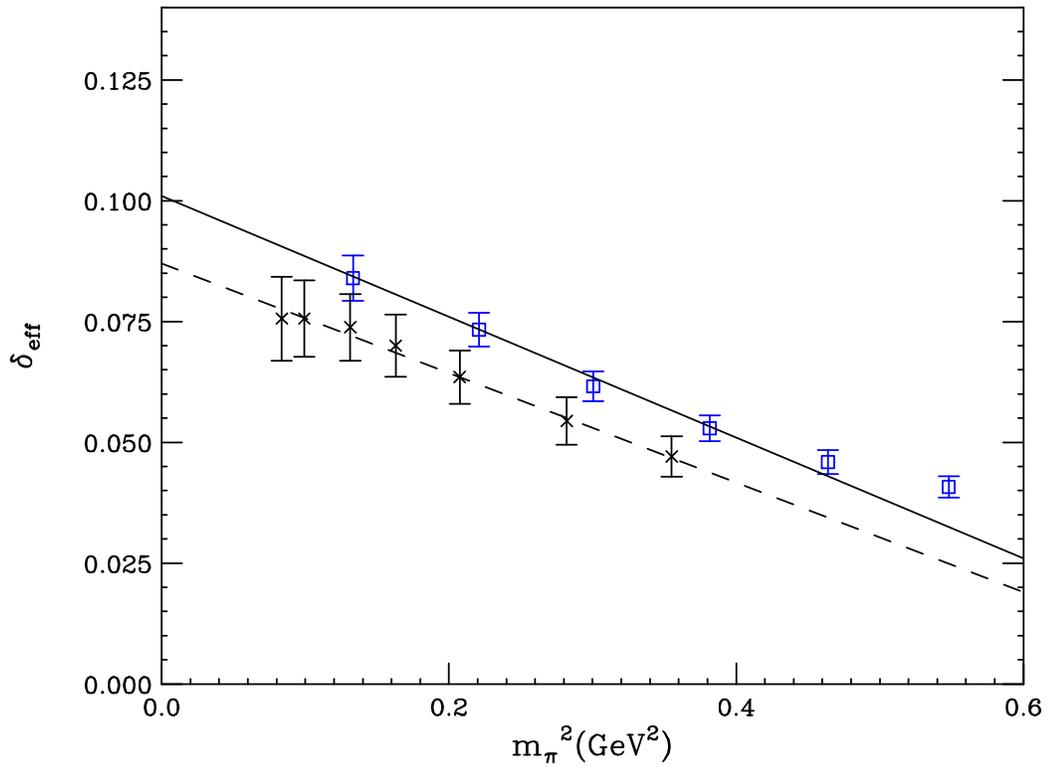}}
\caption[]{ The quenched chiral log parameter $\delta_{eff}$ vs. squared pion mass. Results
from both $\beta=5.7$ ($\times$'s) and 5.9 (boxes) are plotted. The linear fits include 
all mass values for $\beta=5.7$ and the four lightest masses for $\beta=5.9$.}
\label{fig:delta}
\end{figure}

The overall size of quenched chiral loop effects is determined by the parameter $\delta$,
which can be computed from the hairpin insertion mass $m_0$ and the axial vector decay
constant $f_A$ (evaluated in the next Section),
\begin{equation}
\label{eq:delta}
\delta = \frac{m_0^2}{24\pi^2 f_A^2}
\end{equation}
Using the chirally extrapolated values of $m_0$ and $f_A$, we obtain
\begin{equation}
\delta = .099(3), ~~~\beta = 5.7
\end{equation}
and 
\begin{equation}
\delta = .108(4), ~~~\beta = 5.9 .
\end{equation}
It is interesting to consider not only the
value of $\delta$ in the chiral limit, but also the effective
value of $\delta$ at a given quark mass by computing the quantity (\ref{eq:delta}) from
the values of $m_0$ and $f_A$ at that mass. The values of $\delta_{eff}$ vs. pion mass$^2$
for both $\beta=5.7$ and 5.9 are plotted in Fig. 2. This plot shows a rather strong quark mass dependence of the effective QCL parameter, which may provide at least a partial 
explanation of the fact that many
of the determinations of $\delta$ from lattice studies of 
quenched chiral logs \cite{Wittig} have
favored a value of $\delta$ substantially smaller than the phenomenological estimate of
$\delta \approx 0.17$. From Fig. \ref{fig:delta} we see that, for pion masses $>300$ MeV where most studies
have been carried out, the value of $\delta_{eff}$ is smaller than the value in
the chiral limit by as much as a factor two. The decrease of $\delta_{eff}$ with increasing quark mass represents the combined
effect of a decreasing value of $m_0$ and an increasing value of $f_A$ as the quark mass
increases. Although the negative slope of $\delta_{eff}$ has the effect of suppressing
quenched chiral logs, it is nevertheless more consistent to treat $\delta$ as
a constant in fitting to chiral Lagrangian parameters, since the effective mass dependence 
should arise from higher order terms in the chiral expansion.
This is the procedure we adopt in the subsequent analysis of the scalar and pseudoscalar correlators,
where the best $Q\chi PT$ fit favors a value of $\delta$ about half as large as that obtained from the chirally extrapolated
hairpin result. We might expect to find a larger value of $\delta$ if studies were carried out well below
$m_{\pi}=300$ MeV. [It is interesting that a recent study \cite{Kentucky} using overlap fermions, which went as 
low as $m_{\pi}=180$ MeV, found a large value of $\delta$. However, the value $\delta=0.26(3)$ obtained in Ref.
\cite{Kentucky} is much larger than even our chirally extrapolated result of 0.108(4), indicating that there are
other systematic differences in the calculations. Further chiral studies comparing different fermion actions on the
same gauge configurations would be of considerable interest.]

The ``allsource'' quark propagators used to calculate the hairpin correlators \cite{Kuramashi} can also be
used to calculate the topological susceptibility.
Using the integrated anomaly method \cite{SmitVink,chlogsI}, we calculate a winding number for each gauge configuration
from the pseudoscalar charge integrated over the whole lattice. From these winding numbers, we compute
the topological susceptibility $\chi_t=\langle\nu^2\rangle/V$. Using the rho scale, this gives
\begin{eqnarray}
\chi_t & = & (178(4)\;{\rm MeV})^4, ~~~\beta = 5.7 \\
       & = & (171(3)\;{\rm MeV})^4, ~~~\beta = 5.9.
\end{eqnarray}
Results quoted previously \cite{chlogsI} used the charmonium scale, which gives
\begin{eqnarray}
\chi_t & = & (188(4)\;{\rm MeV})^4, ~~~\beta = 5.7 \\
       & = & (190(3)\;{\rm MeV})^4, ~~~\beta = 5.9.
\end{eqnarray}

\section{Pseudoscalar Masses and Decay Constants}

The one-loop chiral Lagrangian analysis of the pseudoscalar and axial-vector
propagators for the $\beta=5.7$ ensemble has been described
previously \cite{chlogsI}. Here we briefly review that analysis and compare
the previously reported results with the new results at $\beta=5.9$. 
We also compare with results of the Alpha collaboration \cite{Wittig,Sommer} 
and discuss some issues associated with extracting the Leutwyler 
parameters $L_5$ and $L_8$, 
or correspondingly $\alpha_5$ and $\alpha_8$.
(Note: Here and elsewhere, we use a conventional notation for the rescaled 
parameters: $\alpha_i = 8(4\pi)^2L_i$).

\begin{figure}
\epsfxsize = 0.80\textwidth
\centerline{\epsfbox{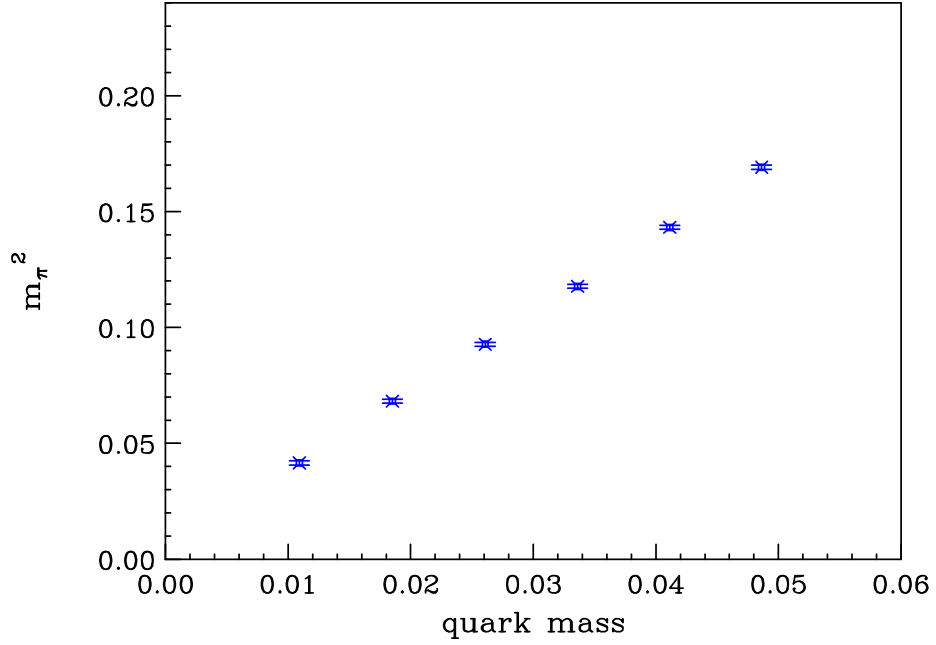}}
\caption{Pion mass squared $m_{\pi}^2$ for equal quark masses for the 
the $\beta = 5.9$ ensemble.}
\label{fig:mdiag}
\end{figure}

\begin{figure}
\epsfxsize = 0.80\textwidth
\centerline{\epsfbox{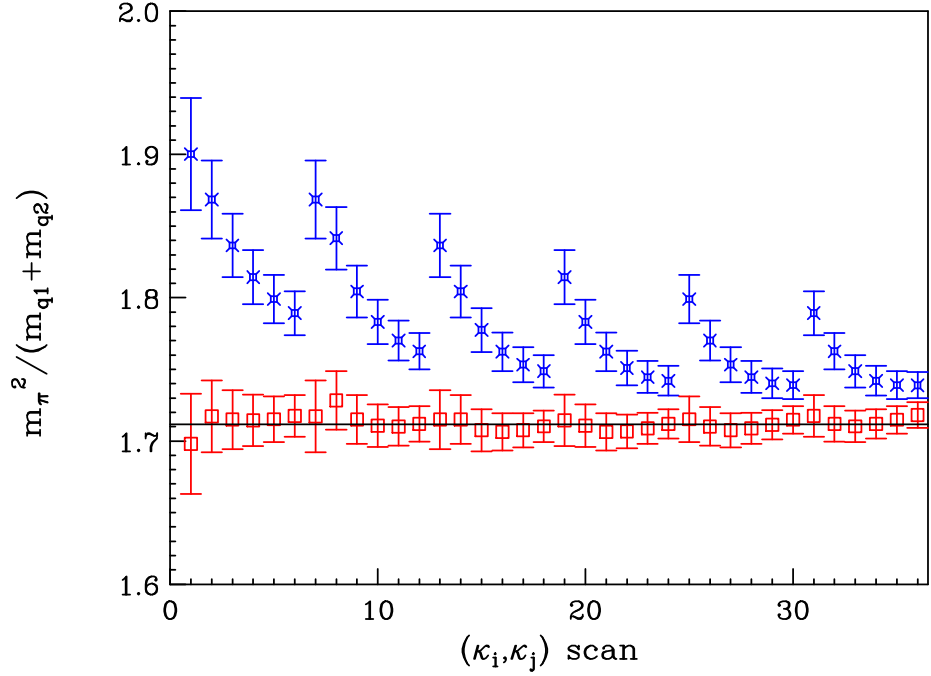}}
\caption[Chiral Slope]{Chiral slope parameter scan for $\beta = 5.9$. 
Individual chiral slopes are denoted by ($\times$).
The scan number is $6*(i-1) +j$
for the pair of $\kappa$ values $(\kappa_i, \kappa_j)$. 
The six $\kappa$ values are ordered from lightest to heaviest quark mass.
Also shown are the ratios to the chiral log fit (boxes) 
which yields a global average of 1.712(39).}
\label{fig:mratio}
\end{figure}

For both ensembles, we calculated propagators using smeared pseudoscalar, local pseudoscalar,
and local axial-vector sources and sinks. The calculations were done for all meson propagators with
both degenerate and nondegenerate quark masses. The chiral Lagrangian parameters were
extracted from a global fit to all pseudoscalar masses and decay constants based 
on one-loop quenched $\chi PT$ for the Lagrangian ${\cal L}_2+{\cal L}_4+{\cal L}_{hp}$, 
as discussed in Section 3. For the lightest pion masses we studied, finite volume effects on chiral
loop integrals are potentially significant, so all one-loop calculations were carried out with
the appropriate finite-volume momentum sums rather than loop integrals. In addition,
quadratic and logarithmically divergent integrals are regularized by subtraction at
a cutoff scale $\Lambda\approx \frac{1}{a}$. To summarize, a generic loop integral
of the form
\begin{equation}
I_{ij} = \frac{1}{\pi^2}\int d^4p\frac{1}{p^2+M_i^2}\frac{1}{p^2+M_j^2}
\end{equation}
is replaced by a cutoff momentum sum 
\begin{equation}
I_{ij} = 16\pi^2\sum_p\left(D(p_i,M_i)D(p_j,M_j)-D(p,\Lambda)^2\right)
\end{equation}
while a quadratically divergent integral
\begin{equation}
I_i = \frac{1}{\pi^2}\int \frac{d^4p}{p^2+M_i^2}
\end{equation}
is replaced by  
\begin{equation}
I_i = 16\pi^2\sum_p\left(D(p,M_i)-D(p,\Lambda)-(\Lambda^2-M_i^2)D(p,\Lambda)^2\right)
\end{equation}
In these expressions, $D(p,M)$ is the free boson propagator and the momentum sums are 
defined according to the physical size of the corresponding lattice volume.
With these definitions, the value of the squared pseudoscalar meson mass
up to first order in $L_5$, $L_8$, and $\delta$ is
\begin{eqnarray}
\label{eq:massformula}
\lefteqn{M^{2}_{ij}}&& =~\chi_{ij} (1+\delta I_{ij})\{1+  
       \frac{1}{f^2}8(2L_8 -L_5)\chi_{ij} [1 
\nonumber \\
   && ~+ \delta (\tilde{I}_{ij} + I_{ij})] 
       +\frac{1}{f^2}8 L_5 \chi_{ij} \delta \tilde{J}_{ij} \} 
\end{eqnarray}
where $r_{0}$ is a slope parameter,
$\chi_i =  2r_{0}m_{i}$,  
$\chi_{ij} = (\chi_i + \chi_j)/2$ and
$m_{i} \equiv \ln(1+ 1/(2\kappa_{i}) - 1/(2\kappa_{c}))$.
Here $\tilde{I}_{ij} = (I_{ii}\chi_i+I_{jj}\chi_j)/\chi_{ij}$, 
\begin{equation}
J_{ij} \equiv \left(I_i+I_j-(M_{ii}^2+M_{jj}^2)I_{ij}\right)/2,
\end{equation}
$\tilde{J}_{ij} = J_{ij}/\chi_{ij}$. 
The loop integrals $I_{ij}$ are defined in \cite{chlogsI}.
The fits for the pseudoscalar masses are shown in 
Figures \ref{fig:mdiag}-\ref{fig:mratio}.

%%%%%%%%%%%%%%%%%%%%%%%%%%%%%%%%%%%%%%%%%%%%%%%%
%%%%%%%%%%%%%%%%%%%%%%%%%%%%%%%%%%%%%%%%%%%%%%%%
%%%%%%%%%%%%%%%%%%%%%%%%%%%%%%%%%%%%%%%%%%%%%%%%

\begin{figure}
\epsfxsize = 0.80\textwidth
\centerline{\epsfbox{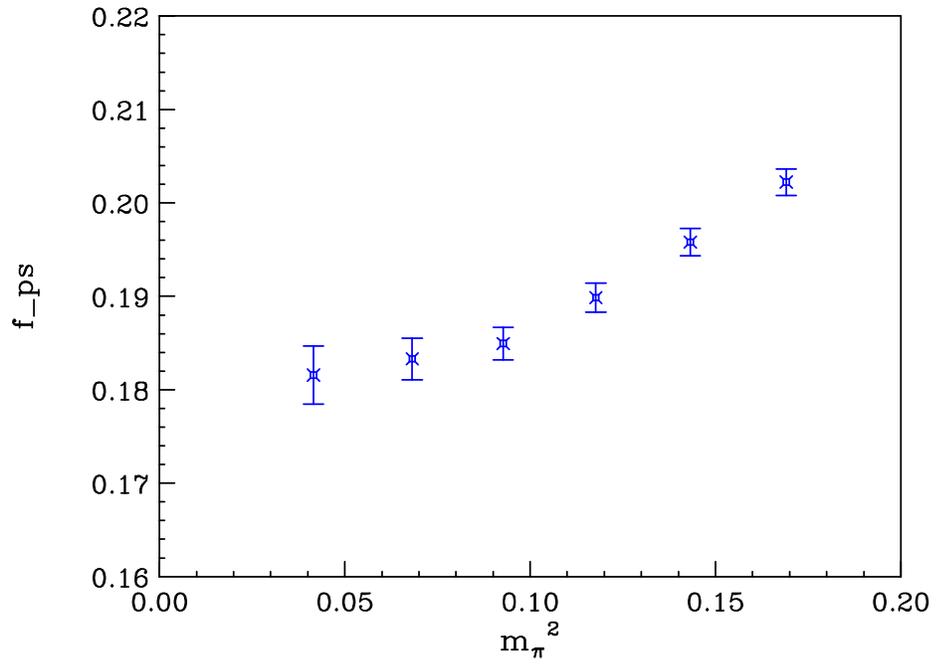}}
\caption{Pseudoscalar decay constant, $f_{\rm ps}$, versus pion mass squared, 
$m_{\pi}^2$ for the $\beta = 5.9$ ensemble. 
Only values for equal quark masses are shown.}
\label{fig:pdiag}
\end{figure}

\begin{figure}
\epsfxsize = 0.80\textwidth
\centerline{\epsfbox{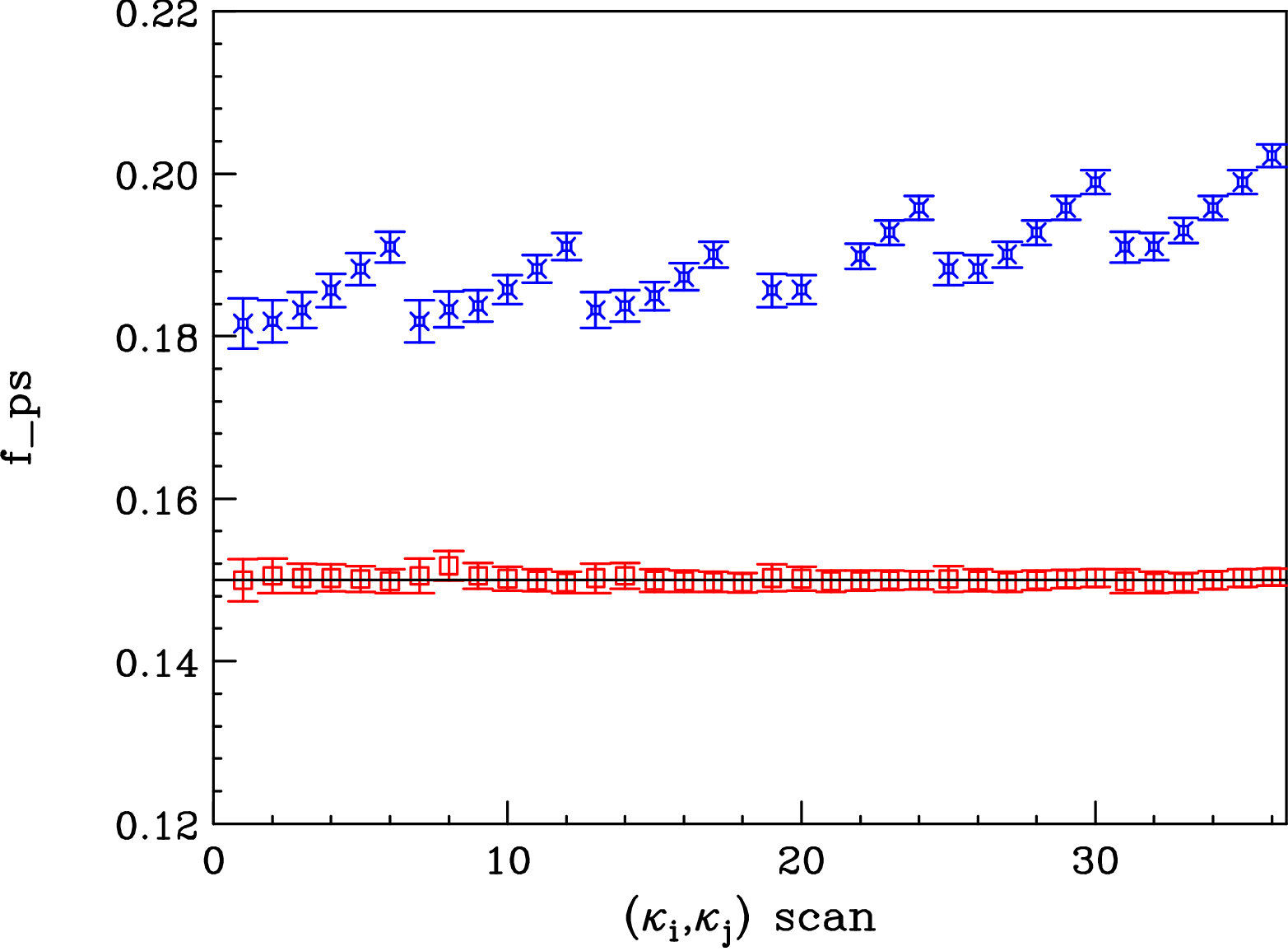}}
\caption{Pseudoscalar decay constant scan for $\beta=5.9$.
Also shown are the ratios to the chiral log fit
which yields a global average $f_{\rm ps} = 0.1501(77)$.
The $(\kappa_i, \kappa_j)$ values are ordered as in Fig.\ref{fig:mratio}.}
\label{fig:pratio}
\end{figure}

For the pseudoscalar decay constants:
\begin{eqnarray}
\label{eq:fpi}
\lefteqn{f_{P;ij}}&&=~\sqrt{2}fr_{0}(1+ 0.25\delta(I_{ii}+I_{jj}+2I_{ij}))\{1
\nonumber \\
  &&~+~\frac{4}{f^2}(4L_8 - L_5)\chi_{ij}[1+\delta(\tilde{I}_{ij} + I_{ij})]
\nonumber \\
  &&~-~\frac{4}{f^2}L_{5}\delta \chi_{ij} [(\frac{\tilde{I}_{ij}}{2} 
          + I_{ij}) - (\tilde{J}_{ii}+\tilde{J}_{jj})]\} 
\end{eqnarray}
The fits for the pseudoscalar decay constants are shown in 
Figures \ref{fig:pdiag}-\ref{fig:pratio}.

\begin{figure}
\epsfxsize = 0.80\textwidth
\centerline{\epsfbox{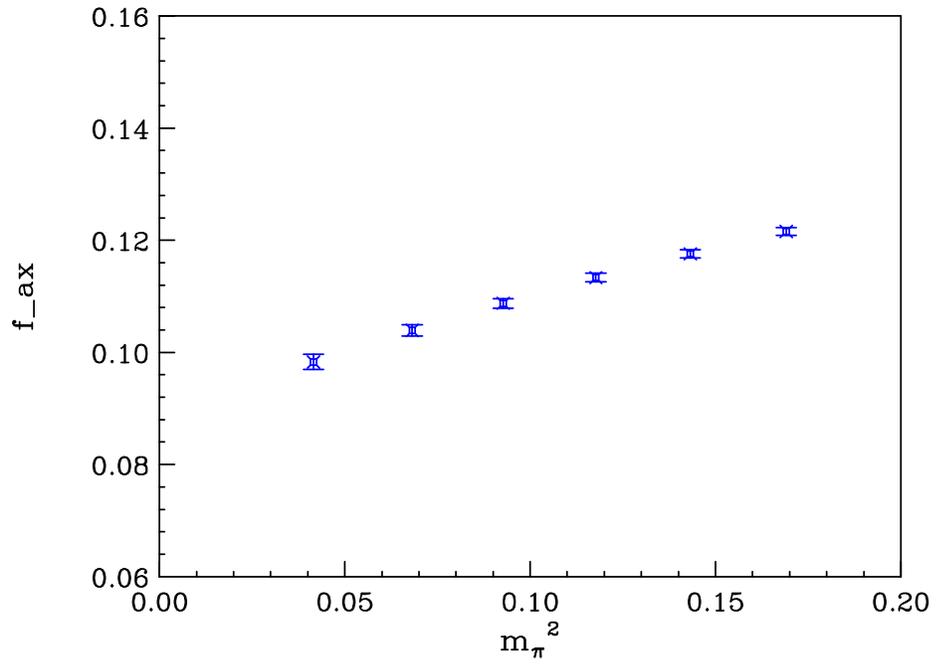}}
\caption{Axial vector decay constant, $f_{\rm ax}$, versus pion mass squared, 
$m_{\pi}^2$ for the $\beta=5.9$ ensemble. 
Only values for equal quark masses are shown.}
\label{fig:adiag}
\end{figure}

\begin{figure}
\epsfxsize = 0.80\textwidth
\centerline{\epsfbox{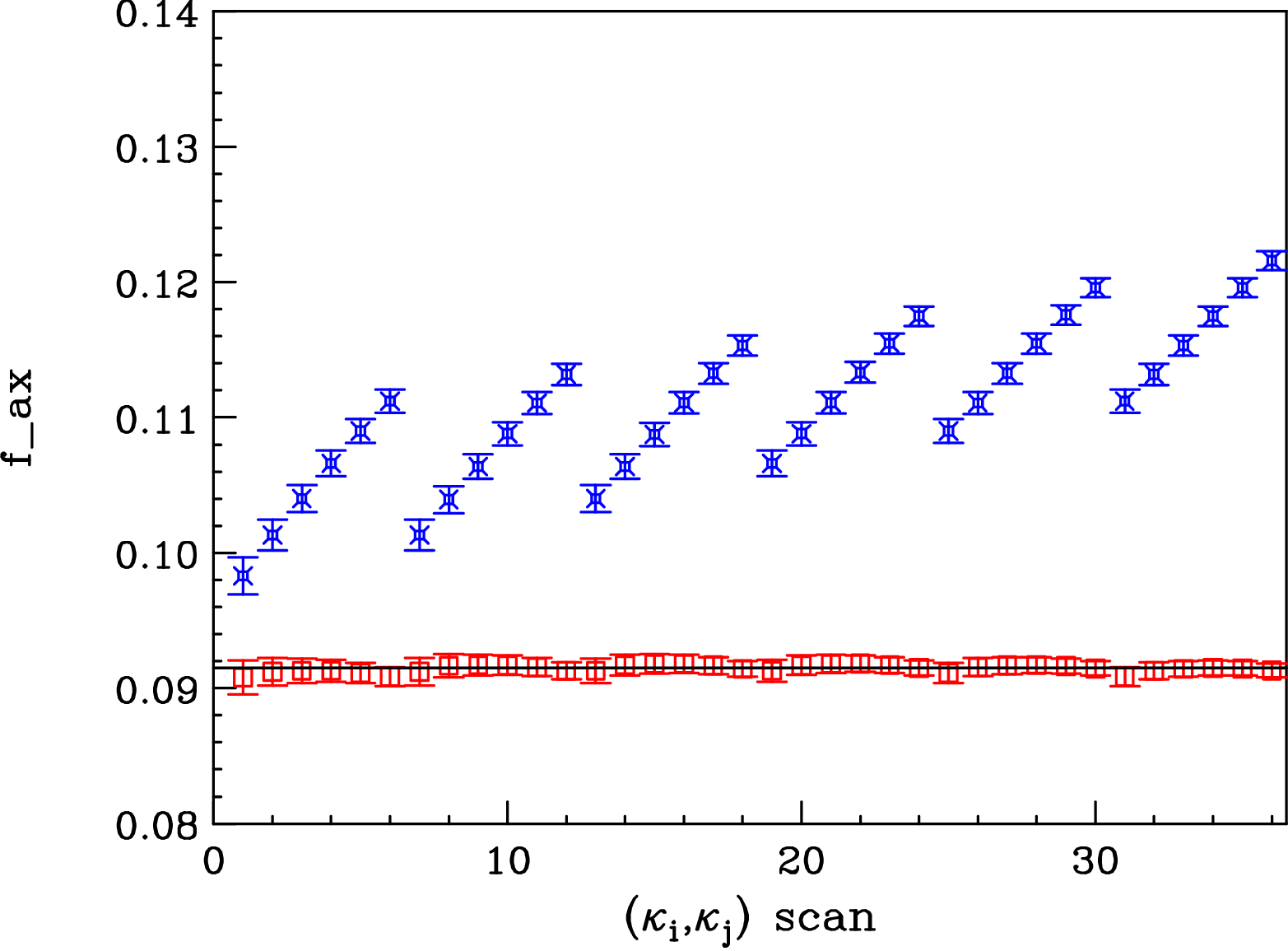}}
\caption{Axial vector decay constant scan, $f_{\rm ax}$ for $\beta=5.9$.
Also shown are the ratios to the chiral log fit 
which yields a global average $f_{\rm ax} = 0.0915(13)$.
The $(\kappa_i, \kappa_j)$ values are ordered as in Fig.\ref{fig:mratio}.}
\label{fig:aratio}
\end{figure}

For the axial vector decay constants, we have:
\begin{eqnarray}
\label{eq:fax}
\lefteqn{f_{A;ij}}&&~=~ \sqrt{2}f(1+ 0.25\delta(I_{ii}+I_{jj}-2I_{ij}))
\nonumber \\
     &&~\{~ 1 + \frac{4}{f^2}\chi_{ij} L_{5}[1 + \delta(2I_{ij} - \tilde{I}_{ij})] \} 
\end{eqnarray}
The fits for the axialvector decay constants are shown in 
Figures \ref{fig:adiag}-\ref{fig:aratio}.
In Figures \ref{fig:mratio}, \ref{fig:pratio}, and \ref{fig:aratio} we show 
both the measured values and the ratio values 
where the fitted chiral log factors in Eqs. \ref{eq:massformula}, \ref{eq:fpi}
and \ref{eq:fax} have been divided out. The results are 
scanned over all combinations of the quark mass values. 

%%%%%%%%%%%%%%%%%%%%%%%%%%%%%%%%%%%%%%%%%%%%%%%%
%%%%%%%%%%%%%%%%%%%%%%%%%%%%%%%%%%%%%%%%%%%%%%%%
%%%%%%%%%%%%%%%%%%%%%%%%%%%%%%%%%%%%%%%%%%%%%%%%
%%%%%%%%%%%%%%%%%%%%%%%%%%%%%%%%%%%%%%%%%%%%%%%%
%%%%%%%%%%%%%%%%%%%%%%%%%%%%%%%%%%%%%%%%%%%%%%%%
%%%%%%%%%%%%%%%%%%%%%%%%%%%%%%%%%%%%%%%%%%%%%%%%

The chiral Lagrangian parameters listed in Table \ref{tab:chparm} are 
the result of a global correlated
fit of the $\chi PT$ expressions to masses and decay constants with 
both equal and unequal quark masses. 
The results obtained from this global fit are generally consistent with
evaluations extracted from more limited fits to equal quark mass data. In particular,
the value of $L_5$ may be estimated directly from the axial vector decay constant $f_A$.
For equal quark masses $f_A$ has no quenched chiral logs, 
so the mass dependence should be linear in the chiral limit,
\begin{equation}
f_A = A + Bm_{\pi}^2
\end{equation}
(Note that, in our notation, $f_A$ includes a factor $\sqrt{2}$ relative to $f_{\pi}$, i.e. it's physical
value is 132 MeV.)
A determination of $L_5$ is obtained from the product of slope and intercept:
\begin{equation}
\label{eq:L5}
L_5 = \frac{A\cdot B}{8}
\end{equation}
Note that $A$ has units of $mass$ and $B$ has units of $(mass)^{-1}$, 
so that $L_5$ is dimensionless and can be evaluated without reference to 
a mass scale. We obtain the results (in GeV using the rho scale)
\begin{equation}
f_A = (.1653(17)+.117(5)m_{\pi}^2)\times Z_A\;\;{\rm GeV}, ~~~\beta = 5.7,
\end{equation}
and
\begin{equation}
f_A = (.1515(20)+.106(7)m_{\pi}^2)\times Z_A\;\;{\rm GeV},  ~~~\beta = 5.9.
\end{equation}
These fits are shown in Figs. \ref{fig:mratio}, \ref{fig:pratio} and \ref{fig:aratio}.
Using $Z_A=0.845$ and $0.865$ for $\beta=5.7$ and $5.9$, respectively, we find
\begin{equation}
\label{eq:L51a}
L_5 (10^3) = 1.72(11), ~~~\alpha_5  =  2.18(14), ~~~\beta = 5.7 
\end{equation}
and
\begin{equation}
\label{eq:L52a}
L_5 (10^3) = 1.50(9), ~~~\alpha_5 = 1.89(11), ~~~\beta = 5.9
\end{equation}
These values are consistent with the global fit results 
in Table \ref{tab:chparm}.

A recent analysis of quenched data from the Alpha collaboration\cite{Sommer} 
obtained a value of $\alpha_5=0.99(6)$. 
Their analysis of the $f_A$ ratios, $R_F(x)$, neglects a factor of 
\begin{equation}
\label{eq:corrS}
 1\over (1+y_{\rm ref} \frac{1}{2}\alpha_5) 
\end{equation}
(c.f Eq. (3.7)-(3.9) of \cite{Sommer})
reflecting the difference between $\chi{\rm PT}$ expansion about zero mass and 
their use of a large reference mass.  Including this factor increases their estimate 
of $\alpha_5$ to
\begin{equation}
\alpha_5 = 1.18(7).
\end{equation}
If only statistical errors are considered, this estimate is still significantly 
below
our values.

%%%%%%%%%%%%%%%%%%%%%%%%%%%%%%%%%%%%%%%%%%%%%%%%
\begin{table}
\centering
\caption{Best fit for the chiral Lagrangian parameters for the $\beta = 5.7$ and~~
$\beta = 5.9$ lattices.}
\vspace*{0.5cm}
\label{tab:chparm}
\begin{tabular}{|c|c|c|}
\hline
 Parameter                &  $\beta = 5.9$     &  $\beta = 5.7$      \\
\hline
 $f=f_{\pi}$              &  $0.091(2)$  &  $0.100(2)$  \\
 $L_5 \times 10^3$        &  $1.55(27)$  &  $1.78(35)$  \\
 $(L_8-L_5/2)\times 10^3$ &  $0.02(5)$   &  $0.14(7)$   \\
 $r_0$                    &  $1.71(8)$   &  $1.99(12)$  \\
 $\delta$                 &  $0.053(13)$ &  $0.059(15)$ \\
\hline
\end{tabular}
\end{table}
%%%%%%%%%%%%%%%%%%%%%%%%%%%%%%%%%%%%%%%%%%%%%%%%

We can identify a number of differences in the two analyses 
which could account for
much of the discrepancy.  First, our data uses somewhat coarser lattices 
and an independent determination of the clover coefficient. 
This could affect the comparison
with the Alpha analysis although they see little lattice spacing dependence
in their data.
There are also differences in methodology. 
Our results are obtained from a linear fit to $f_A(m_{\pi}^2)$
where the value of $L_5$ is determined from the dimensionless 
product of the slope and the intercept of the fit.  This method is
insensitive to physical scale parameter but is quite sensitive to the 
value of the axial-vector renormalization constant, 
being proportional to $Z_A^2$. 
In the context of the tadpole renormalization scheme \cite{LM} our results in 
Eqs. \ref{eq:L51a}-\ref{eq:L52a} include the perturbative renormalization 
constants $Z_A = 0.845$ and $Z_A = 0.865$ at $\beta = 5.7$ and $\beta = 5.9$, 
respectively.  These values are derived from the perturbative formulas 
given in Ref. \cite{Gupta}.

As an alternative method, we could determine the renormalization 
factors by requiring a fit to the physical value of $f_{\pi} = 93$ MeV.  
The result will now be 
sensitive to the choice of scale parameter. Fixing the scale with the rho mass, 
the renormalization constants are given by $Z_A=0.80$ for $\beta = 5.7$ and
$Z_A = 0.87$ for $\beta = 5.9$. Using these renormalization constants modifies
our predictions to 
\begin{eqnarray}
 L_5 (10^3) = 1.54(10)    &,& \alpha_5 = 1.95(13), ~~~\beta = 5.7 \\
 L_5 (10^3) = 1.52(9)     &,& \alpha_5 = 1.91(11), ~~~\beta = 5.9 
\end{eqnarray}
This result is still substantially above the value obtained in 
the Alpha analysis.  We note that, in our results,  the mass dependence 
of the tadpole renormalization factor contributes about 17\% to the slope 
of $f_A(m_{\pi}^2)$.
For linear extrapolations of $f_A(m_{\pi}^2)$ the above method is actually
exactly equivalent to the Alpha ratio method with the inclusion of the 
correction factor (Eq. \ref{eq:corrS}).  

In the Alpha procedure, the physical value of $f_{\pi}$ is used along 
with a particular choice of scale parameter.  
(This procedure, which involves the physical value of 
$f_{\pi}$, and/or $m_{\rho}$, is somewhat
adhoc since the quenched values for $f_{\pi}$ and $m_{\rho}$ need not
agree precisely with the Particle Data Group.)
We note that caution is required in any direct comparison between 
the quenched and unquenched values of $\alpha_5$.  We have emphasized 
that there are no quenched
chiral logs which affect the extrapolation of $f_A(m_{\pi}^2)$ for
equal quark masses.  However there is a strong chiral log effect in the
unquenched theory.
The Gasser and Leutwyler analysis\cite{Leut85} of the one-loop chiral logs 
determines $\alpha_5$ from the physical values of $f_K/f_{\pi}$
\begin{equation}
\alpha_5 = 2.8 \pm 0.6 - 3 \ln(\mu/m_{\eta}).
\end{equation} 
Hence, $\alpha_5 = 0.5$ for a large scale, $\mu = 4\pi f_{\pi}$, $\alpha_5 = .99$
for $\mu = 1$ GeV and $\alpha_5 = 1.76$ for $\mu = m_{\rho}$.
The quenched theory really corresponds to the leading order of the $1/N_c$
expansion of the full theory which can not be isolated phenomenologically
from the higher order contributions.

\section{The Quenched Scalar Propagator}

One of the most dramatic effects of quenched chiral loops is observed in the scalar
valence propagator. The behavior of this propagator for the $\beta=5.7$
ensemble was studied in \cite{scalar}. There it was found that the propagator was
well-described as a combination of a short-range positive exponential associated
with a heavy ($>1$ GeV) scalar $\bar{q}q$ meson state and a long-range {\it negative} tail arising
from the $\eta'$-$\pi$ intermediate state. In the quenched approximation, the 
$\eta'$-$\pi$ loop diagram exhibits not just a quenched chiral logarithm, but
a quenched chiral {\it power}, with infrared behavior $\sim d^4p/p^6$. This 
long-range component is well-described in both shape and 
magnitude by the finite-volume $\eta'$-$\pi$ loop
calculation. [Note: In the $\beta=5.7$ results, 
the $\approx 2$ standard deviation discrepency between the
data and the $\chi PT$ calculation for $t>7$ (c.f. Fig. 10 of Ref. \cite{scalar})
we now believe to be a statistical fluctuation. See below.]
As discussed in \cite{scalar} the one-loop term that dominates the
scalar propagator at small quark mass is determined with no adjustable parameters
by the chiral Lagrangian parameters 
$m_{\pi}, f,$ and $r_0$, already fixed from the analysis of
the pseudoscalar sector. In particular, the long-range scalar propagator exhibits
a very strong mass dependence in the light quark regime, 
which is very well explained by the dependence of
the finite volume one loop contribution on $m_{\pi}^2$. 

\begin{figure}
\epsfxsize = 0.80\textwidth
\centerline{\epsfbox{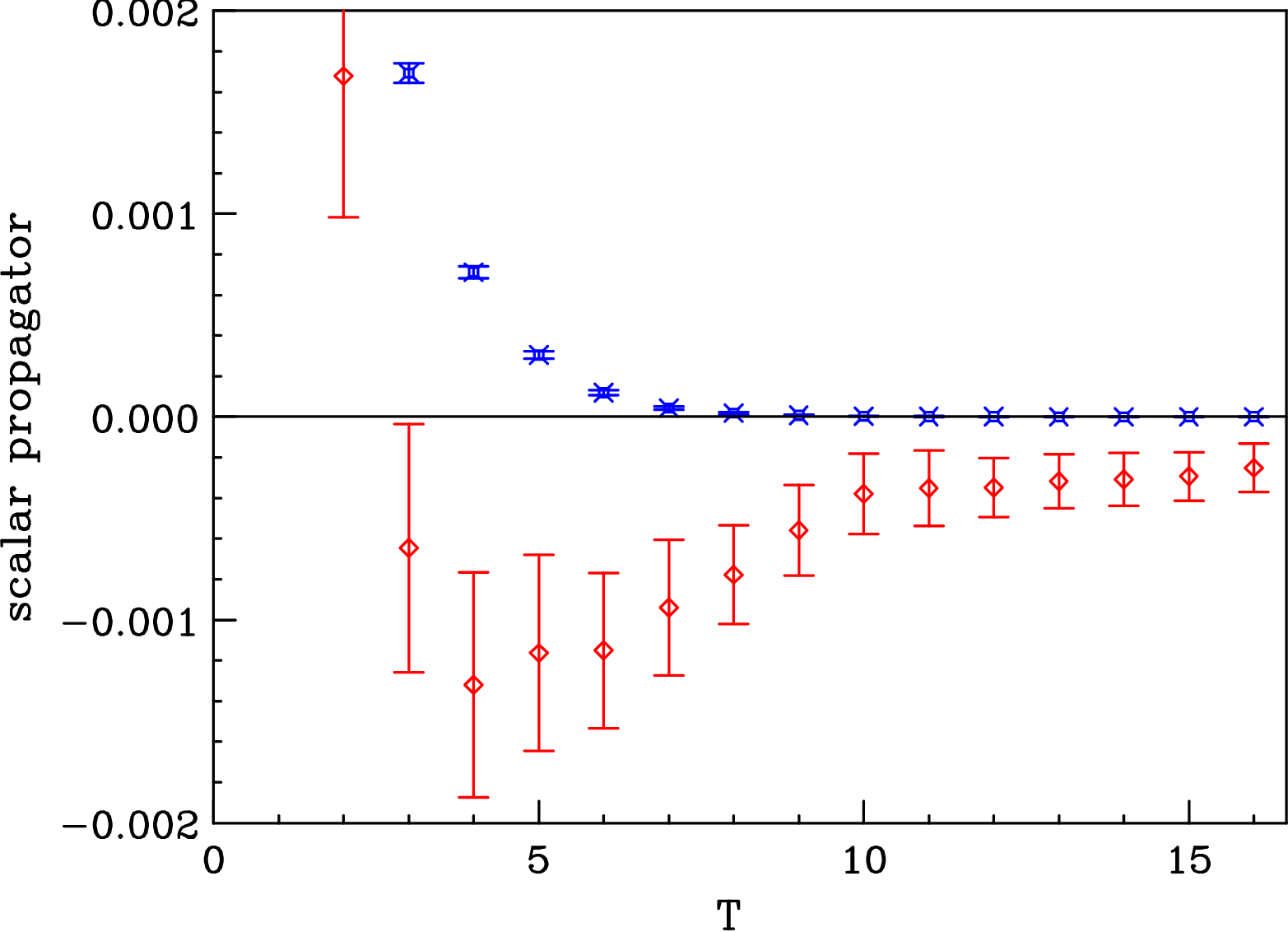}}
\caption{Isovector scalar correlator for the $\beta = 5.9$ $16^3\times 32$ lattices.
The lightest  
($\kappa_{q} = \kappa_{\bar q} = .1397$) (red $\diamond$) and heaviest
($\kappa_{q} = \kappa_{\bar q} = .1382$) (blue $x$) correlators are shown.}
\label{fig:negscalar}
\end{figure}

\begin{figure}
\epsfxsize = 0.65\textwidth
\centerline{\epsfbox{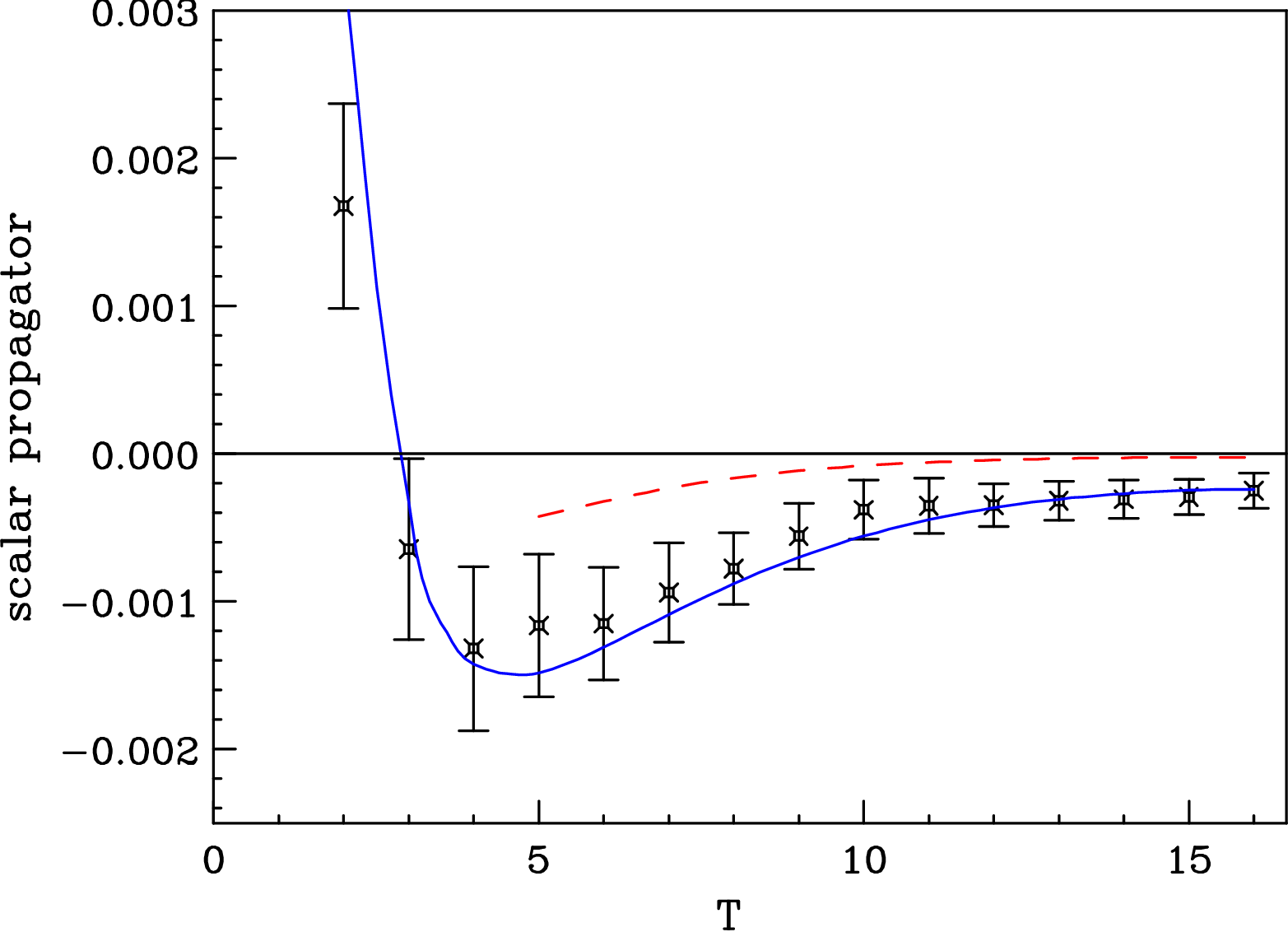}}
\caption[]{The scalar correlator for $\beta=5.9$ and $\kappa=.1397$. 
The solid line is a fit consisting of the sum of a scalar meson pole 
and an $\eta'$-$\pi$ loop diagram. The dashed line is the loop term
in the infinite volume limit.}
\label{fig:scprop}
\end{figure}

The analysis of the scalar correlator for $\beta=5.9$ confirms   
the main conclusions of Ref. \cite{scalar}. (See Figures
\ref{fig:negscalar}-\ref{fig:scprop})
For the $5.9$ ensemble, we find excellent agreement
between the long-range behavior of the correlator and the $Q\chi PT$ loop calculation (which contains
no adjustable parameters), with the theoretical formula matching the data all the way out
to the largest time separation of $t=16$.
%%%%%%%%%%%%%%%%%%%%%%%%%%%%%%%%%%%%%%%%%%%%%%%%%%%%%%%%%%%%%%%%%%%%%%%%
The details of the analysis of the $\beta=5.7$ data were discussed in \cite{scalar}.
The $\beta=5.9$ ensemble was analyzed similarly, though with some simplifications. First,
in \cite{scalar} both smeared and local sources were used, and a 
factorization procedure allowed an estimate of the mass of the excited 
$a_0^*$ meson. 
This produced a good fit to the scalar correlator
all the way down to t=1. For the analysis of the 5.9 data, we have 
not carried out an independent estimate of the excited $a_0^*$ 
contribution, but have simply rescaled the parameters obtained from
the 5.7 analysis, holding this contribution fixed in the fits which 
determine the ground state $a_0$ parameters. 
For the 5.9 data we used a time range $t \geq 3$, so the fits 
were insensitive to the choice of excited state parameters.
The other difference in the 5.9 analysis is that the lightest
pion mass (330 MeV) is somewhat heavier than that for the 5.7 study (275 MeV). In \cite{scalar}
the formula used to parametrize the $\eta'$-$\pi$ contribution was obtained by resumming 
iterated bubble graphs to all orders. In the analysis of the 5.9 data, we have found that only
the one-loop graph is significant, so we have discarded higher order bubbles in the formula
used to fit the correlator. Thus, defining the time-dependent zero-momentum scalar correlator as
\begin{equation}
\Delta(t) \equiv \sum_{\vec{x}}\langle\bar{\psi}_1\psi_2(\vec{x},t)\bar{\psi}_2\psi_1(0)\rangle
\end{equation}
we fit the lattice correlator to the function (c.f. Eq. (14) of Ref. \cite{scalar})
\begin{equation}
\label{eq:scalarfit}
\Delta(t) \sim 32r_0^2\frac{f_s^2}{2m_s}e^{-m_st} + 4r_0^2\tilde{B}_{hp}(t)
\end{equation}
where $\tilde{B}(t)$ is the $\vec{p}=0$ Fourier transform of the $\eta'$-$\pi$ bubble
graph, calculated in a finite volume:
\begin{equation}
B_{hp}(p) = \frac{1}{VT}\sum_k\frac{1}{[(k+p)^2+m_{\pi}^2]}\frac{-m_0^2}{(k^2+m_{\pi}^2)^2}
\end{equation}
Note that $m_{\pi}$, $r_0$, and $m_0$ are already determined from the pseudoscalar correlator analysis,
so the only two fit parameters are the scalar mass and decay constant $m_s$ and $f_s$. 
In Fig. \ref{fig:scprop} we show the data for the scalar
correlator at $\beta=5.9$ and the lightest quark mass, $\kappa=.1397$, 
along with a fit given by the function (\ref{eq:scalarfit}). 

%%%%%%%%%%%%%%%%%%%%%%%%%%%%%%%%%%%%%%%%%%%%%%%%%%%%%%%%%%%%%%%%%%%%%%%%
%%%%%%%%%%%%%%%%%%%%%%%%%%%%%%%%%%%%%%%%%%%%%%%%%%%%%%%%%%%%%%%%%%%%%%%%

\begin{figure}
\epsfxsize = 0.65\textwidth
\centerline{\epsfbox{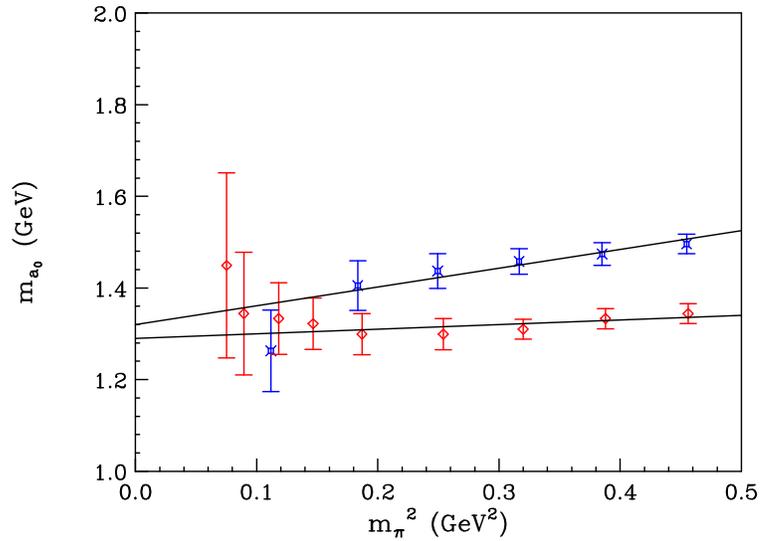}}
\caption[]{Mass of the $a_0$ meson for $\beta=5.7$ (boxes) and $\beta=5.9$ ($\times$'s).}
\label{fig:a0mass}
\end{figure}

The mass of the $a_0$ meson obtained from our fits 
is plotted in Fig. \ref{fig:a0mass}. In the chiral limit 
we obtain (using the rho scale)
\begin{eqnarray}
m_{a_0} & = & 1285(60)\;MeV\;\;{\rm at}\;\beta=5.7 \\
        & = & 1326(86)\;MeV\;\;{\rm at}\;\beta=5.9
\end{eqnarray}
%%%%%%%%%%%%%%%%%%%%%%%%%%%%%%%%%%%%%%%%%%%%%%%%%%%%%%%%%%%%%%%%%%%%%%%%
%%%%%%%%%%%%%%%%%%%%%%%%%%%%%%%%%%%%%%%%%%%%%%%%%%%%%%%%%%%%%%%%%%%%%%%%
%%%%%%%%%%%%%%%%%%%%%%%%%%%%%%%%%%%%%%%%%%%%%%%%%%%%%%%%%%%%%%%%%%%%%%%%
\begin{figure}
\epsfxsize = 0.65\textwidth
\centerline{\epsfbox{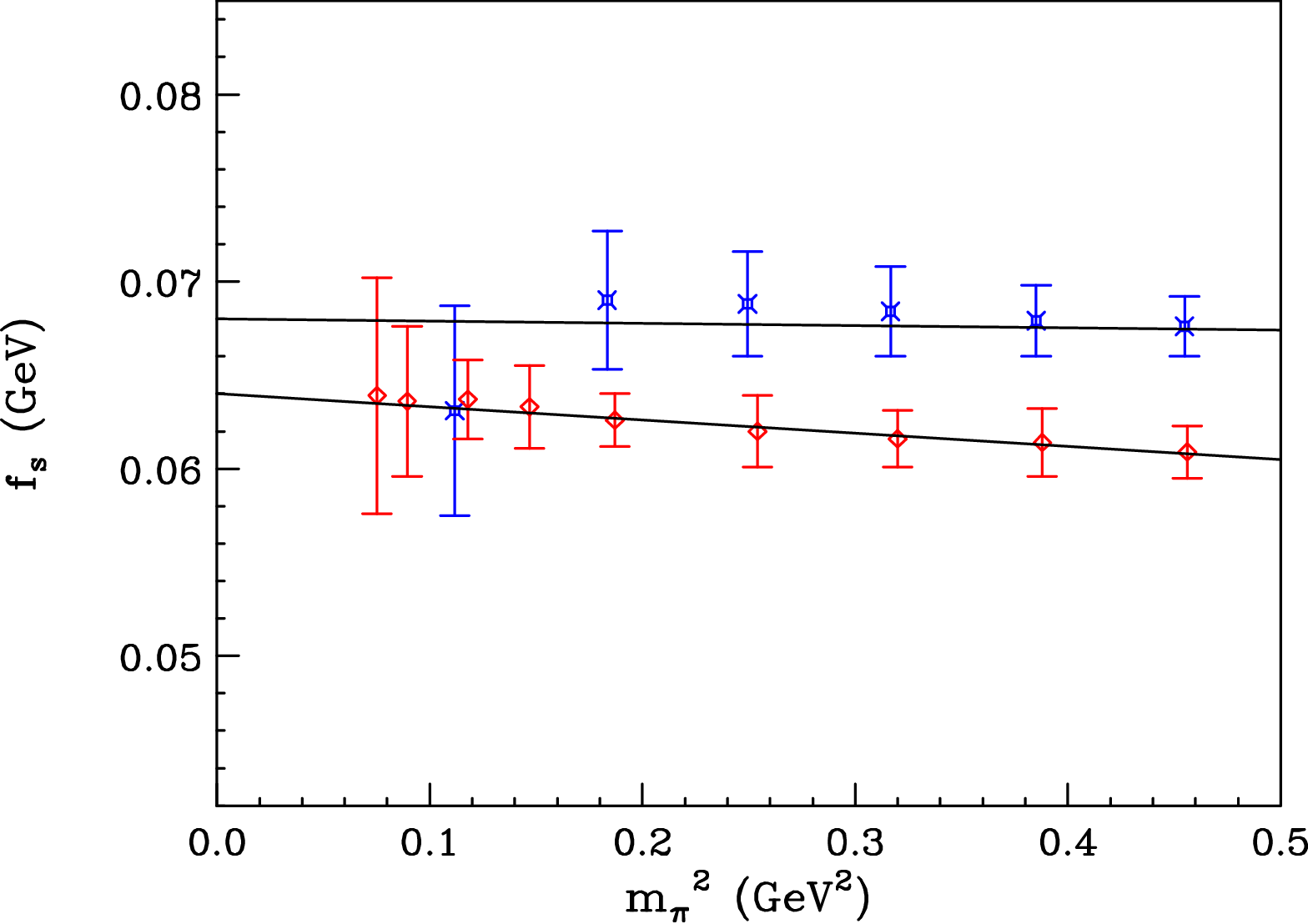}}
\caption[]{The scalar decay constant of the $a_0$ meson 
for $\beta=5.7$ (boxes) and $\beta=5.9$ ($\times$'s).}
\label{fig:a0fs}
\end{figure}
%%%%%%%%%%%%%%%%%%%%%%%%%%%%%%%%%%%%%%%%%%%%%%%%%%%%%%%%%%%%%%%%%%%%%%%%

In the chiral loop calculation of the $\eta'$-$\pi$ intermediate state 
contribution, we have included
the effect of finite physical volume, replacing momentum integrals by 
discrete sums. For the lightest pion masses
the finite volume effect is quite large, due to the infrared behavior of 
the loop contribution. It is interesting to note that the Monte Carlo results 
for the scalar correlator agree quite nicely with the prediction of
{\it finite volume} Q$\chi$PT, but disagree significantly with the 
corresponding infinite volume calculation. This strong finite volume effect is
also seen by the RBC collaboration\cite{Prelovsek}.
The dashed line in Fig. \ref{fig:scprop} is the infinite volume loop 
calculation. 
For $t>5$, the correlator is dominated by
the $\eta'$-$\pi$ loop, and it is clear that the Monte Carlo result 
exhibits the predicted enhancement from the finite
volume effect. From the point of view of Dirac eigenmodes, 
the finite volume dependence predicted by
$\chi$PT arises from a subtle interplay between exact zero modes 
and near zero modes. In the infinite volume limit, the
contribution of exact zero modes vanishes, but for finite volume these 
modes are essential for reproducing the correct
chiral behavior. Since the MQA pole shifting ansatz consists of 
repositioning some exact zero modes, it is
reassuring to see that the resulting correlators exhibit excellent 
agreement with the finite volume effect
predicted by Q$\chi$PT.

Finally, the scalar decay constant, $f_s$, of the $a_0$ meson obtained from our fits 
is plotted in Fig. \ref{fig:a0fs}. In the chiral limit 
we obtain (using the rho scale)
\begin{eqnarray}
f_s     & = & 64(5)\;MeV\;\;{\rm at}\;\beta=5.7 \\
        & = & 68(3)\;MeV\;\;{\rm at}\;\beta=5.9
\end{eqnarray}

%%%%%%%%%%%%%%%%%%%%%%%%%%%%%%%%%%%%%%%%%%%%%%%%%%%%%%%%%%%%%%%%%%%%%%%%
 
\section{Conclusions}

The lattice calculations described in this paper and our previous works \cite{chlogsI, scalar, lat02} have focused on the chiral properties of meson 
correlators in the flavor singlet and nonsinglet pseudoscalar and nonsinglet 
scalar channels. In addition to exhibiting 
for the first time a number of anomalous chiral effects 
due to quenching, the results have confirmed a level of overall consistency 
of the low energy chiral Lagrangian description of meson properties in QCD, 
and provided quantitative estimates of the relevant chiral Lagrangian 
parameters. 

A central feature of this study is the accurate calculation of the pseudoscalar 
flavor-singlet hairpin insertion responsible for the gluonic component of 
the $\eta'$ mass. In this calculation the MQA pole shifting ansatz 
has a particularly salutary effect. Since the pseudoscalar hairpin 
insertion arises from the $U_A(1)$ anomaly, it is particularly
sensitive to topological features of the gauge field. As a result, the exceptional
configuration problem in the hairpin is even more serious than in the valence correlators.
The MQA procedure allows the first detailed study of the time-dependence
of the hairpin correlator. The most striking feature of this time-dependence is that it 
is quite accurately described {\it at all time separations} by a the simple chiral
Lagrangian diagram consisting of two pion propagators on either side of a 
momentum independent mass insertion. Values obtained for the field 
renormalization parameter $\alpha_{\Phi}$, which parametrizes the momentum 
dependence of the hairpin insertion, are very small and consistent
with zero. Perhaps even more remarkable is the fact that the hairpin correlator 
exhibits an absence of excited state contamination, with the ground-state 
double-pole diagram giving a complete description of the correlator. 
This result is confirmed not only by the time-dependence
of the hairpin correlator, but also by a detailed comparison of hairpin and valence
correlators with both smeared and local sources, as discussed in Ref. \cite{chlogsI}.
Since we know from the valence correlator that the local $\bar{\psi}\gamma^5\psi$ operator 
creates a state which includes a substantial excited state component along with the 
ground state pion, we conclude that these excited states are decoupled from the hairpin
vertex itself. This may be viewed as a plausible extension of the OZI rule. 
Calculation of vector and axial-vector hairpin correlators \cite{IsgurThacker} has 
confirmed that $q\bar{q}$ annihilation in these channels is highly suppressed 
compared to the anomaly-enhanced pseudoscalar hairpin diagram, as expected from 
OZI phenomenology. The absence of evidence for excited states in the pseudoscalar hairpin 
indicates that annihilation from these excited pseudoscalar states is similarly 
OZI suppressed. Only if the $q\bar{q}$ are in a
Goldstone state do they have an unsuppressed pair-annihilation amplitude.
The possibilities for extending these calculations to investigate the details of the 
connection between quark pair creation and annihilation and topological charge 
fluctuations are intriguing.

A global fit to the pseudoscalar masses and decay 
constants obtained from the valence (nonsinglet) correlators
as a function of the quark masses yields an estimate of the quenched chiral 
log parameter $\delta$, the pion decay constant $f_{\pi}$, the slope parameter 
$r_0$, and the Leutwyler parameters $L_5$ and $L_8$ (Table \ref{tab:chparm}). 
We have found that a chiral Lagrangian analysis at the one-loop
level provides a good description of the data over the range of masses studied, 
for both equal and unequal quark mass. The fit values of the 
quenched chiral log parameter $\delta$, $\sim .05-.06$, are rather
small, but consistent with the values obtained directly from the hairpin diagram, 
averaged over the mass region studied. The chirally extrapolated value of 
$\delta$ ($\sim .10$) evaluated from the hairpin correlator is somewhat
larger, but still smaller than the phenomenological estimate of $\sim 0.17$. 
There may be some indication (cf. Fig. 2) that the value of $\delta$ is increasing 
for smaller lattice spacing so that, in the continuum limit, it might be closer 
to the phenomenological estimate. Similarly, the values of 
$L_5$ ($\alpha_5$) we obtain are somewhat larger than the recent results of 
Heitger et.al. \cite{Sommer}. In any case,
since the lattice spacings we have studied are fairly coarse, the overall 
consistency of our fits suggests that, even for finite lattice spacing, the low 
energy dynamics is well-described by a chiral Lagrangian, with the main lattice 
spacing effects consisting of corrections to the parameters. 

The behavior of the nonsinglet scalar propagator at light quark mass provides a 
particularly useful probe of chiral dynamics, and the success of the chiral 
Lagrangian description (Section 6) is impressive.  The prominent $\eta'$-$\pi$ 
loop contribution is completely determined in terms of the chiral
parameters $m_{\pi}$, $r_0$, and $m_0$, extracted from the pseudoscalar 
analysis. The loop calculation agrees well with the lattice data in magnitude, 
time-dependence, and pion mass dependence. 
For the box size ($\sim 2$ fm) and pion masses we have used, the loop diagram 
exhibits a strong finite volume effect, and the agreement with the data is 
only satisfactory if the loop calculation is carried out in a finite volume. 
It would be interesting to carry out the numerical scalar
propagator calculation on different size lattices to explicitly observe this 
finite volume effect.  In view of the agreement we see on a single size box over 
a wide range of pion masses, we would expect
the chiral loop diagram to also give a good description of the finite volume 
dependence, as long as the box is still large compared to the QCD scale. 
The sensitivity of the $\eta'$-$\pi$ loop diagram to finite volume effects 
makes it especially useful for probing the role of zero modes in finite volume 
calculations. For extremely large boxes (large compared to the chiral scale), 
zero modes and global topology should be irrelevant, e.g. ensemble averages 
over any fixed topology should converge to the same result in the infinite 
volume limit. On the other hand, for volumes which are large with respect 
to the QCD scale but still comparable to the chiral scale, the exact zero modes 
contribute in an essential way to the dynamics described by the chiral Lagrangian. 
The techniques developed here can easily be used to study the role of zero 
modes and global topology in finite volume chiral dynamics. 
This can be done, for example, by studying the contributions to the ensemble average 
for various correlators as a function of the global topological charge. We have seen 
that the integrated anomaly method of Smit and Vink \cite{SmitVink}, 
after MQA improvement, 
provides a convenient way of estimating the global topological index $\nu$ of a 
configuration. Calculation of $\nu$ using exactly chiral (e.g. overlap or domain wall) 
fermions would be ideal, but the approach used here based on clover improved Wilson 
fermions is more economical and appears to be quite effective for studying 
issues which do not depend crucially on having exact integer valued $\nu$'s. 
For example, the results for the topological susceptibility (Section 4) 
are in good agreement with other estimates. 
It would be interesting to carry out a more detailed investigation of the scalar and 
pseudoscalar correlators studied here, decomposing the ensemble averages into various
topological charge sectors. This would provide useful insight into the role of global 
topology in chiral dynamics at finite volume.

After properly accounting for quenched chiral effects in the scalar propagator,
the $a_0$ masses (at $\kappa_c$) are extracted. 
The results, $m_{a_0} = 1.33(5)$ and $1.34(6)$ GeV for $\beta = 5.9$
and $5.7$ respectively are considerably larger than the observed $a_0(980)$ resonance
mass.
The value for $m_{a0} = 1.33(5)$ GeV suggests that there is a large effect
when internal quark loops are included or that the
observed $a_0(980)$ resonance is a distinct state, possibily a $K\bar{K}$ ``molecule'' 
(which would not appear in the quenched approximation),
and not an ordinary $q\bar{q}$ meson.

%%%%%%%%%%%%%%%%%%%%%%%%%%%%%%%%%%%%%%%%%%%%%%%%%%%%%%%%%%%%%%%%%%%%%%%%

\section*{Acknowledgements}

The work of W. Bardeen and E. Eichten was performed 
at the Fermi National Accelerator Laboratory, which is 
operated by University Research Association,
Inc., under contract DE-AC02-76CHO3000. 
The work of H. Thacker was supported in part by the
 Department of Energy under grant DE-FG02-97ER41027.

%%%%%%%%%%%%%%%%%%%%%%%%%%%%%%%%%%%%%%%%%%%%%%%%%%%%%%%%%%%%%%%%%%%%%%%%

\end{document}